\documentclass[twocolumn,showpacs,amsmath,amssymb]{revtex4}
\usepackage{graphicx}
\usepackage{color}
\usepackage{graphicx,psfrag,booktabs}

\begin{document}

\title{Novel oscillation for the indirect coupling between magnetic nanoparticles}
\author{C. H. Chang and T. M. Hong}
\affiliation{Department of Physics, National Tsing Hua University, Hsinchu 300, Taiwan, Republic of China}
\date{\today}

\begin{abstract}
We study the prospect of using magnetic nanoparticles for the diluted magnetic seminconductors.
The Ruderman-Kittel-Kasuya-Yosida formula is modified by explicitly taking into account the role of charge carriers inside the nanoparticles in addition to those in the medium. Calculations are done analytically. The final form of the coupling between nanoparticles is similar to the original formula except for additional phase terms which render a novel oscillatory feature with respect to the particle size, an interesting analogy to the same behavior when we vary their distance. This is to be contrasted to the previous approach which did not include the inner carriers and only favored a ferromagnetic coupling. 
 The effect of inevitable deviations from a perfect sphere is estimated by the Born approximation. Our derivations can be readily generalized to finite temperatures.
\end{abstract}


\maketitle

\section{INTRODUCTION}
DMS (Diluted Magnetic Semiconductors) have been hailed as a potential spintronics device
to integrate the computing power of CPU and the storage ability of the hard disc\cite{alan,dietl}. By intimately coupling them, we hope to come up with a device that can compute while recording, and vice versa. The way to prepare the DMS sample includes implanting transition-metal ions in a semiconducting template or using the digital doping.
When the dosage increases, the ferromagnetism mediated and favored in the dilute limit by the charge carriers appears. Although annealing was found to increase the Curie temperature by forcing out the carriers originally trapped in the antisites, the optimum $T_c$ is still only around 110K. There have been reports of room-temperature DMS, but they were often plagued by the appearance of magnetic clusters\cite{clustering}. This renders them unusable because the coupling between the electronics and magnetism we set out to preserve is lost. 

In this paper, we discuss the scenario\cite{super} of utilyzing the magnetic nanoparticles for DMS, whose movement is hindered by their size and is therefore less inclined to clustering. In the mean time, they are small enough to enter the superparamagnetic regime which allows them to carry a net moment but without a permanent magnetizaion. Unlike the pestering clusters that appeared unwantedly, the size and shape of these nanoparticles and their spacing can be controlled before we cover them by a semiconducting layer. Similar proposal has been studied in the context of Giant Magnetoresistance\cite{toro} for a metallic medium. Note that, although the underlying mechanism is expected to follow the theory by Ruderman, Kittel, Kasuya, and Yosida (RKKY), how the actual coupling between nanoparticles differs from the RKKY formula for point impurities is not clear. Previous workers assumed the validity of the original formula and concentrated on the mathematical difficulty of integrating over spins in a sphere\cite{genkin, summation}, wire, slab, or semi-infinite plane\cite{slab}. Discrepancies\cite{qiang} were found between their results and the experimental data\cite{qiang, lopez, toro, du}. We believe the root of this inconsistency lies in their failure to acknowledge the mediating role of the charge carriers $\it {inside}$ the nanoparticles. This neglect is particularly fatal when the medium is semiconducting because, while the previous approach would start from a ferromagnetic coupling for $\it {all}$ pairs of spins, ours encompasses the possibility of its becoming antiferromagnetic when the pair distance varies by as little as a quarter of $5\sim 10$A - typical Fermi wavelength for metals. Most interestingly, our improvement predicts an oscillatory feature for this indirect coupling with respect to their size, in complement to the same behavior to their separation.

When a long-range magnetic order is established among the nanoparticles, energy levels of all charge carriers will be split by the Zeeman energy. Since the internal field is much stronger inside the nanoparticle, we shall neglect the splitting in the medium for the time being. However, it can be easily reinstated for the self-consistent calculation of the transition temperature for such based DMS.  One fact that comes in handy is that these nanoparticles shall be sparsely spaced in the DMS, namely, their separation $R$ is much greater than their size $a$. This allows us to take the asymptotic limit of the special functions involved in matching the incoming, reflected, and transmitted eigenfunctions across the boundary defined by the nanoparticle. In the previous approach\cite{genkin, skomski, summation}, the effective coupling term after integrating over all pairs of spins looks the same as the original RKKY formula: $J(R_{12}) \vec{S}_{\rm {eff,1}} \cdot \vec{S}_{\rm {eff,2}}$ with
\begin{equation}
\lim_{R_{12}\gg a}J(R_{12})=J_{0}\frac{\cos(2k_{F}R_{12})}{(2k_{F}R_{12})^{3}}\\
\end{equation}
where $J_{0}$ is a parameter of the system, $k_{F}$ is the Fermi momentum of the medium, and
$\vec{S}_{\rm{eff,i}}$ is the effective magnetic moment for the i-th nanoparticle. The oscillation with respect to $2k_{F}R$ remains valid.
It is not expected that this oscillatory and decaying behavior should be totally anulled when we include extra charge carriers in the nanoparticles. A naive guess is to replace the $2k_F R_{12}$ by $2k_F (R_{12}-2a)+2k_{F,n}(2a)$ where $k_{F,n}$ is the Fermi momentum of the nanoparticles. Detailed derivations later, after considering the Zeeman energy, show that a more appropriate form is to lump the effect of these new messengers in additional phase terms as:
\begin{equation}
J(R_{12})=J_{0}\frac{\cos(2k_{F}R_{12}+\phi_{1}+\phi_{2})}{(2k_{F}R_{12})^{3}}
\label{eq:coupling}
\end{equation}
where $\phi_{i}$ denotes the phase shift due to the different dispersion inside the $i$-th nanoparticle.

\section{DERIVATIONS AND RESULTS}
\begin{figure}[h!]
\psfrag{a}{$a$}
 \psfrag{C}{$k_{F\uparrow}$}
 \psfrag{D}{$k_{F\downarrow}$}
\psfrag{E}{$k^{'}_{F}$}
\includegraphics[width=0.4\textwidth]{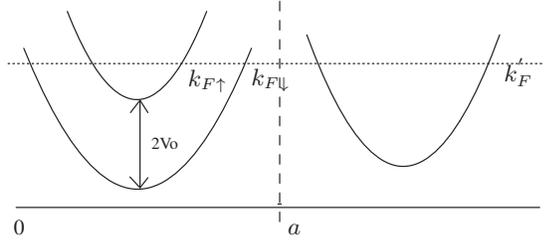}
\caption{\small The energy bands on the left belong to the charge carriers inside the magnetic nanoparticles. They are split by the Zeeman energy, $V_{0}$. In contrast, only one Fermi momentum $k'_{F}$ exists in the medium on the right.}
\label{fig:kf}
\end{figure}

Due to the internal field, energy bands inside the nanoparticle are split by twice the Zeeman energy $V_{0}$, as shown schematically in Fig.(\ref{fig:kf}), and different Fermi momentum $k_{F\sigma}$ can be defined for each spin $\sigma$. 

\subsection{Wave functions of free electrons}
Let us first calculate the indirect coupling between a spherical magnetic nanoparticle and a single impurity spin.
The general form of the electron eigenfunction is easiest expressed in spherical coordinates with the origin set at its center. Symmetry allows us to separate the variables into
\begin{equation}
\psi_{lm\sigma}=S_{l\sigma}(r)Y_{lm}(\theta,\phi)
\label{eq:psi}
\end{equation}
where the radial part consists of the spherical Bessel functions, $j_{l}$ and $n_{l}$. Since the latter diverges at the origin, we have to exclude it for $r\le a$:
\begin{equation}
S_{l\sigma}(r)=
\begin{cases}
j_{l}(k_{\sigma}r),  &\text { if $r\le a$};\\
A_{l\sigma}j_{l}(k^\prime r)+B_{l\sigma}n_{l}(k^\prime r), &\text{ if $r\ge a$}
\end{cases}
\end{equation}
where the three momenta are related by
\[\frac{k^{\prime 2}}{2m}=\frac{k_{\uparrow}^{2}}{2m}+V_{o}=\frac{k_{\downarrow}^{2}}{2m}-V_{o}\]
The coefficients, $A_{l\rho}$ and $B_{l\rho}$, can be determined by matching the boundrary condition as:
 \begin{align}
 \notag
&A_{l\sigma}(k^\prime )=(k^\prime a)^{2}\left[\frac{k_{\sigma}}{k^\prime}j_{l+1}(k_{\sigma}a)n_{l}(k^\prime a)-j_{l}(k_{\sigma}a)n_{l+1}(k^\prime a)\right]\\
\notag
&B_{l\sigma}(k^\prime )=(k^\prime a)^{2}\left[ j_{l}(k_{\sigma}a)j_{l+1}(k^\prime a)-\frac{k_{\sigma}}{k^\prime}j_{l+1}(k_{\sigma}a)j_{l}(k^\prime a)\right]
\end{align}
Finally, after properly normalizing the eigenfunction in Eq.(\ref{eq:psi}), we denote it by $\Psi_{k^\prime lm\rho}$.

\subsection{Magnetic oscillation}
\begin{figure}
\psfrag{a}{$a/\lambda_{F}$} \psfrag{E}{$E/(J_{0}*10^{-11})$}
\includegraphics[width=0.4\textwidth]{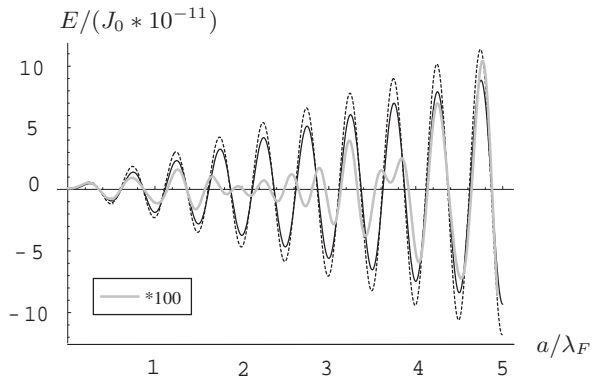}
\caption{\small The coupling energy is calculated via Eq.(\ref{eq:rho2}) by choosing the Zeeman splitting, $V_0/E_{\rm{F}}$, to be 0.001 and 0.1 (represented by dark and gray solid lines respectively). It can be seen that, as $V_0$ decreases, the calculated value approaches that of the previous approach\cite{genkin, skomski,summation} (in dashed line) which neglected the carriers inside the nanoparticle but carried out the spherical integration over spins. The separation $R$ is set at $30 \lambda_{F}$.}
\label{fig:match}
\end{figure}

Spatial variation of the magnetization can be calculated from the
eigenfunction obtained above:
\begin{equation}
\langle M(\overrightarrow{R})\rangle=\sum_{k'<k'_{F}}\sum_{lm}\big[\Psi^{*}_{k'lm\uparrow}\Psi_{k'lm\uparrow}-\Psi^{*}_{k'lm\downarrow}\Psi_{k'lm\downarrow}\big]
\label{eq:firstM}
\end{equation}
Since the nanoparticle is assumed to be spherical, the azimuthal dependence drops out for $M$.
Furthermore, the fact that  
$R\gg a$ in DMS allows us to Taylor expand and retain just
the lowest order term in $a/R$:
\begin{equation}
\langle M({R})\rangle \simeq Q(k^\prime_{F})\frac{\sin(2k^\prime_{F}{R})}{4{R}^{3}}
+W(k^\prime_{F})\frac{\cos(2k^\prime_{F} {R})}{4{R}^{3}}
\label{eq:rho}
\end{equation}
where $Q(k^\prime_{F})$ and $W(k^\prime_{F})$ are constants that depend on the parameters $(k^\prime_{F},a,V_{o})$:
\begin{align}
\notag
Q(k^\prime_{F})&=\Big[\frac{-A_{l\uparrow}^{*}A_{l\uparrow}+B_{l\uparrow}^{*}B_{l\uparrow}}{N_{l\uparrow}}-\frac{-A_{l\downarrow}^{*}A_{l\downarrow}+B_{l\downarrow}^{*}B_{l\downarrow}}{N_{l\downarrow}}\Big]\\
W(k^\prime_{F})&=\Big[\frac{A_{l\uparrow}^{*}B_{l\uparrow}+A_{l\uparrow}B_{l\uparrow}^{*}}{N_{l\uparrow}}-\frac{A_{l\downarrow}^{*}B_{l\downarrow}+A_{l\downarrow}B_{l\downarrow}^{*}}{N_{l\downarrow}}\Big] .
\end{align}
The normalization $N_{l\sigma}\equiv 1/\sqrt{A_{l\sigma}^2+B_{l\sigma}^2}$ and $A_{l\sigma}$, $B_{l\sigma}$, are to be evaluated at $k^\prime =k^\prime_F$.
In order to compare with the original RKKY formula, we
rewrite Eq.(\ref{eq:rho}) as:

\begin{equation}
\langle M(R)\rangle\approx 2\sqrt{Q^{2}+W^{2}}\cdot \frac{\cos(2k_{F}^\prime R+\phi(k'_{F}))}{(2k^\prime_{F}R)^{3}}
\label{eq:rho2}
\end{equation}
where $\phi=-\arctan(Q/W)$. Fig.\ref{fig:match} demonstrates that, when we set the dispersion relation of the nanoparticles to be the same as in the medium and gradually turn off the Zeeman splitting, the energy obtained from Eq.(\ref{eq:rho2})  becomes indistinguishable from that of the previous work\cite{genkin,skomski,summation}. This proves that the action of matching the boundary condition is equivalent to carrying out the integration over spins inside the nanoparticle. Therefore, the coefficient, $2\sqrt{Q^{2}+W^{2}}$, in Eq.(\ref{eq:rho2}) can be thought of
as an effective magnetic moment for the nanoparticle. Note that the phase term $\phi$ is due to different dispersions relation for these two regions of charge carriers and was absent in the previous approach\cite{genkin,skomski,summation}.

\begin{figure}[h!]
\psfrag{a}{$a/\lambda_{F}$}
\psfrag{S_{ef}}{$S_{\rm {eff}}$}
\includegraphics[width=0.4\textwidth]{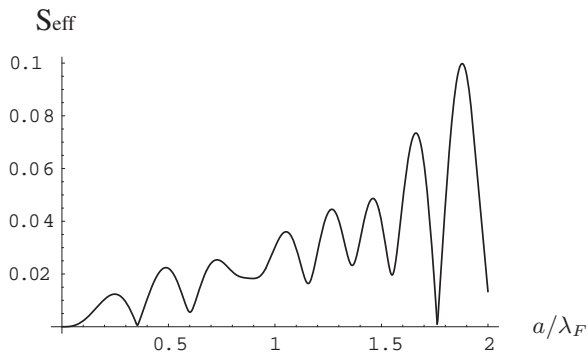}
\caption{\small $S_{\rm {eff}}$ oscillates with $a/\lambda_{F}$ where
$\lambda_{F}={2\pi}/{k'_{F}}$ is the Fermi wavelength characteristic of the
medium.}
\label{fig:eff}
\end{figure}

\begin{figure}[h!]
\psfrag{a}{$a/\lambda_{F}$} \psfrag{b}{$\phi$ $(-\pi/2\sim\pi/2)$}
\includegraphics[width=0.4\textwidth]{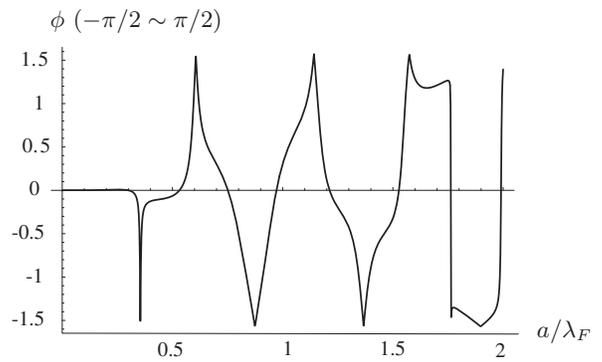}
\caption{\small The extra phase term, $\phi$, due to the different dispersion of the nanoparticle is shown to exhibit the same oscillatory behavior as Fig.\ref{fig:eff} when the particle size increases.}
\label{fig:phase}
\end{figure}

The effective moment, $S_{\rm {eff}}$, is plotted against the size of the nanoparticle in Fig.\ref{fig:eff}. The oscillatory feature exists for all densities of the medium carriers. This is unlike the previous prediction\cite{genkin, skomski, summation}, which favors the ferromagnetic coupling in the dilute limit according to the original RKKY formula.
Similar oscillation is also found in the phase term in Fig.\ref{fig:phase}. Expectedly, this phase is only appreciable when the particle size exceeds a quarter of the Fermi wavelength, $\lambda_{F}$, to allow for a possible change of sign for the coupling.

\subsection{Correction due to a nonspherical surface}
\begin{figure}[h!]
\psfrag{a}{$a/\lambda_{F}$}\psfrag{E}{$E/(J_{0}*10^{-9})$}
\includegraphics[width=0.4\textwidth]{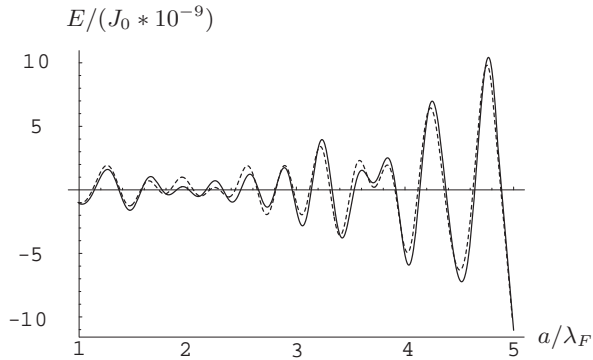}
\caption{\small The energy of the coupling term is plotted as a function of the mean radius of the particle. The distance between the nanoparticle and the impurity spin is set at $R=30 \lambda_{F}$. The dashed line is for a perfect sphere, while the solid line is for a nanoparticle characterised by the surface, $r(\theta,\varphi)=0.03\lambda_{F} \times \sin6\theta$.}
\label{fig:roughcoupling}
\end{figure}

Real samples are never a perfect sphere. We estimate the effect of this deviation or roughness by treating it as a scattering potential.
\begin{figure}[h!]
\psfrag{A}{$r=f(\theta,\varphi)$}
\psfrag{B}{$\psi_{\vec{k},\sigma}(\vec{r})$}
\psfrag{C}{$\phi_{\sigma}(\vec{R})$} \psfrag{D}{$\vec{R}$}
\includegraphics[width=0.4\textwidth]{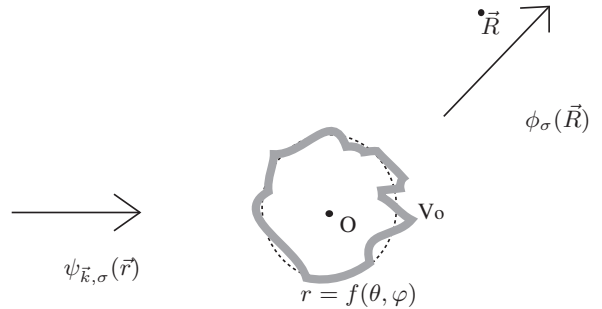}
\caption{\small Schematic plot of a matter wave scattered by the irregularity, or the potential due to the deviation from a perfect sphere, of a nanoparticle. This process is spin-dependent.
Wavefunction $\psi_{\vec{k},\sigma}(\vec{r})$ denotes the incoming wave,
$\phi_{\sigma}(\vec{R})$ the outgoing wave, and $\vec{R}$ is the position of the observer.}
\label{fig:scatter}
\end{figure}
In Fig.\ref{fig:scatter}, an incoming wave is shown schematically to be scattered by the potential $U$ due to the deviation of the boundary, $r=f(\theta,\varphi)$, of a realistic nanoparticle from a perfect sphere. When observed at a distance $|\overrightarrow{R}|$
much longer than the size of the potential, the scattered wave can be estimated by the Born
approximation in the scattering theory:
\begin{equation}
\phi_{\sigma}(\vec{R})\approx \psi_{\vec{k},\sigma}(\vec{r})-2m\int
d^{3}\vec{r} G_{\sigma}(\vec{r},\vec{R}) U_{\sigma}(\vec{r})
\psi_{\vec{k},\sigma}(\vec{r})
\label{eq:last}
\end{equation}
where $m$ is the effective mass of carriers in the medium, $\psi_{\vec{k}, \sigma}(\vec{r})$ is the incoming wave for spin $\sigma$ taken from  Eq.(\ref{eq:psi}), and $\phi_{\sigma}(\vec{R})$ the
wavefunction after the perturbation. The scattering potential
$U_{\sigma}$ comes from and is therefore defined by the deviation of the real rough boundary from a perfect sphere:
\begin{equation}
U_{\uparrow}(\vec{r})=-U_{\downarrow}(\vec{r})=V_0\cdot\big[\Theta(f(\theta,\varphi)-r)-\Theta(a-r)\big]
\end{equation}
where $V_0$ is the Zeeman energy and $\Theta(r)$ denotes the unit step function. The unperturbed wavefunction is taken to be Eq.(\ref{eq:psi}), with which we can construct the Green's function
$G_{\sigma}(\vec{r},\vec{R})$.
After substituting the unperturbed wavefunction by the  $\phi$ in Eq.(\ref{eq:last}), the effect of an irregular surface on the magnetic coupling can be estimated.
For comparison, the size dependence of the coupling energy similar to Fig.\ref{fig:match} is plotted in Fig.\ref{fig:roughcoupling} for both a perfect sphere and a rough surface characterized by $r=0.03\lambda_{F} \times \sin6\theta$. It shows that not only the coupling strength receives a quantitative change, but the nodes of zero coupling get shifted. 

\subsection{Magnetic interaction between nanoparticles}
\begin{figure}[h!]
\psfrag{B}{$(S_{\rm{eff,1}},\phi_{1})$} \psfrag{C}{$(S_{\rm{eff,2}},\phi_{2})$}
\includegraphics[width=0.4\textwidth]{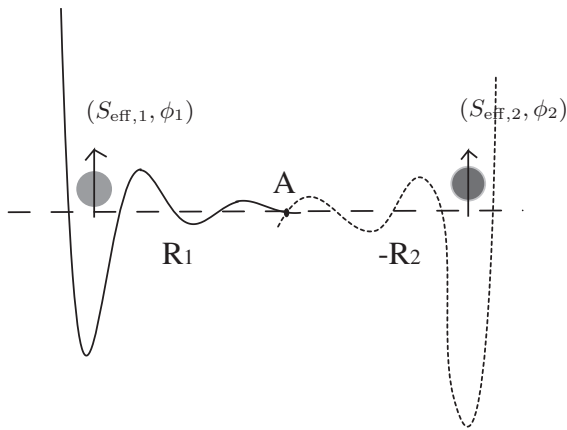}
\caption{\small Imagine a fictitious impurity spin at position A. Since we already know how each nanoparticle is coupled to the single spin, the net coupling between the nanoparticles can be constructed by the concept of interference.}
\label{fig:interact}
\end{figure}

Finally, we are ready to generalize the above procedures to between two nanoparticles. Let us resort to physical arguments rather than matching the boundary condition at now two spherical surfaces, which no longer exhibits the rotational symmetry and therefore does not warrant the separation of variables. Imagine a virtual point spin sitting between the two nanoparticles. We already know how each nanoparticle interacts with the spin, which predicts a polarized Fermi sea like a tidal wave in Fig.\ref{fig:interact} according to Eq.(\ref{eq:rho2}). Two tidal waves are generated by these two nanoparticles and meet to create interference patterns. Irrespective of where the virtual spin sits, we expect the best coupling to occur when the waves are in phase\cite{bruno}. Namely, when the difference between $(2k^\prime_{F}R_{1}+\phi_{1})$ and $(2k^\prime_{F}R_{2}+\phi_{2})$ equals integer multiples of $2\pi$. While the effective moment $S_{\rm{eff,i}}$ and 
phase term $\phi_{i}$ can be estimated independently in Eq.(\ref{eq:rho2}), we therefore expect the coupling energy between two nanoparticles to exhibit the form, $E=J({R}_{12})\vec{S}_{\rm{eff,1}}\cdot \vec{S}_{\rm{eff,2}}$ with
\begin{equation}
J(\textbf{R}_{12})\simeq J_{o}
\frac{ \cos (2k^\prime_{F}{R}_{12}+\phi_{1}+\phi_{2})}{(2k^\prime_{F}{R}_{12})^{3}.}
\end{equation}


\section{SUMMARY and DISCUSSIONS}

In this article, we calculated the indirect coupling between sparsely spaced magnetic
nanoparticles in conductors. Instead of quoting the original RKKY form for point spins, we rederive the formula by replacing the plane waves by the more suitable wavefunction which takes into account that the charge carriers inside the nanoparticles may have a different dispersion relation from those in the medium. We show that the matching of boundary condition for the wavefunction is equivalent to performing the spherical integration over the spins inside the nanoparticle except for an extra phase term. This phase term is crucial at rendering an oscillatory dependence for the coupling on the particle size. This is an interesting analogy to the same behavior on their separation. Our improvement is particularly relevant if the nanoparticles are to be utilized at manufacturing a new breed of DMS, because without the phase term the indirect coupling will mistakenly start with a ferromagnetic coupling for all spins. Although our discussions focus on spherical nanoparticles, formalisms are provided on how to deal with irregular surfaces. Our derivations follow the same procedures as the standard RKKY model and, therefore, can be as easily generalized to finite temperatures.

\begin{figure}[h!]
\psfrag{a}{$a/\lambda_{F}$}\psfrag{E}{$E/(J_{0}*10^{-7})$}
\includegraphics[width=0.4\textwidth]{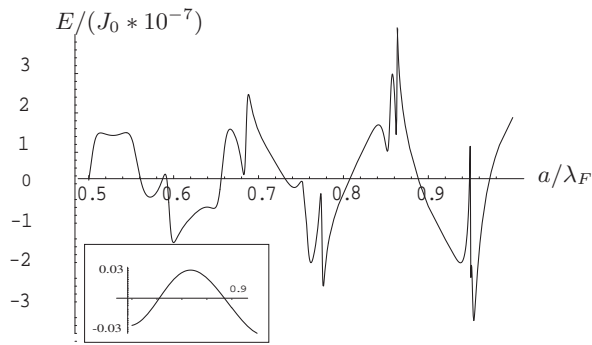}
\caption{\small Coupling energy between a nanoparticle and a point impurity is plotted as a function of the particle size. In contrast to Fig.\ref{fig:match}, part of which is shown in the inset, the period shortens when the effective mass of the carriers in the medium is larger (four times than those in the nanoparticle for this figure). We set the distance at $R=20\lambda_{F}$ and the Zeeman splitting $V_0$ equals one tenth of the
Fermi energy.}
\label{fig:effectivemass}
\end{figure}

For simplicity, we have used the same dispersion relation for the carriers in the nanoparticle as those in the medium, except for the Zeeman splitting. There are two comments here. Firstly, we checked the effect of a different effective mass for the nanoparticle in Fig.\ref{fig:effectivemass}. It shows that, as the inner carriers get heavier, the oscillation with respect to the particle size becomes less periodic and with a shorter wavelength. Secondly, we found that the same procedures in subsection IIC were also able to reproduce the dashed line in Fig.\ref{fig:match}. Instead of using the deviation from a perfect sphere, a Zeeman splitting (energy $\pm V_0$ for up and down spins respectively) within the radius $a$ serves as the scattering potential. Apparently, there will be no effect whatsoever, should $V_0$ equal zero. So the previous results\cite{genkin, skomski, summation} are reproduced at the first order of the Born approximation, as has been done by Pogorelov and coworkers\cite{born}. The phase term that is unique to our approach can only be obtained when we go to higher-order terms.

We benifit from discussions with Professors C. R. Chang, H. H. Lin, and D. W. Wang. Support by the National Science Council in Taiwan under grants 95-2112-M007-046-MY3 and 95-2120-M007-008 is acknowledged.



\end{document}